\documentstyle[11pt]{article}
\newcommand{\be}{\begin{equation}}
\newcommand{\ee}{\end{equation}}
\newcommand{\bea}{\begin{eqnarray}}
\newcommand{\eea}{\end{eqnarray}}

\begin{document}

 \title{\bf A New Graded Algebra Structure on Differential Polynomials: Level Grading and its Application to the Classification of Scalar Evolution Equations in $1+1$ Dimension}
\author {
              {\bf  Eti M\.IZRAH\.I}\\
        {\it  Department of Mathematics, Istanbul Technical University}\\
   {\it  Istanbul, Turkey}\\
        {\it  e-mail: mizrahi1@itu.edu.tr}\\ {} \\
        {\bf  Ay\c{s}e H\"{u}meyra B\.ILGE}\\
        {\it  Faculty of Engineering and  Science, Kadir Has University}\\
        {\it  Istanbul, Turkey}\\
        {\it  e-mail: ayse.bilge@khas.edu.tr}\\ {} \\
          }\maketitle

\begin{abstract}
\baselineskip=12pt \noindent We define a new grading, that we
call the ``level grading", on the algebra of polynomials
generated by the derivatives $u_{k+i}=\partial^{k+i}u/\partial
x^{k+i}$ over the ring $K^{(k)}$ of $C^{\infty}$ functions of
$u,u_1,\dots,u_k$.  This grading has the property that the
total derivative  and  the  integration by
parts with respect to $x$ are filtered algebra maps. In addition, if $u$ satisfies
an evolution equation $u_t=F[u]$ and $F$ is a level homogeneous
differential polynomial, then the total derivative with respect
to $t$, $D_t$,  is also  a filtered algebra map. Furthermore if $\rho$ is  level homogeneous over
$K^{(k)}$, then the top level part of $D_t\rho$ depends on $u_k$ only.
This property allows to
determine the dependency of $F[u]$ on $u_k$ from the top level
part of the conserved density conditions. We apply this
structure to the classification of ``level homogeneous" scalar
evolution equations  and we obtain the top level parts of
integrable evolution equations of ``KdV-type", admitting an
unbroken sequence of conserved densities at orders
$m=5,7,9,11,13,15$.
\end{abstract}
\baselineskip=16pt

\section{ Introduction}

The classification of evolution equations has been a long
standing problem in the literature on evolution equations. The
existence of higher symmetries, an infinite sequence of
conserved densities, of a recursion operator, the Painleve
property of reduced equations have been proposed as
integrability tests. Among these, we follow the ``formal
symmetry" method of Mikhailov-Shabat-Sokolov (MSS)
\cite{MSS91}, which is based on the remark that if the
evolution equation admits a recursion operator than, its
expansion as a pseudo-differential should satisfy an operator
equation.  The solvability of the coefficients of this
pseudo-differential operator in the class of local functions
necessitates that certain quantities be conserved. These
quantities are called
 ``canonical densities" and their existence is proposed as an integrability test \cite{MSS91}.

In \cite{MSS91}, a  preliminary classification of third order
equations has been obtained by the formal symmetry method.  The
classification given in \cite{MSS91} asserts the existence of a
three classes of candidates  for evolution equations, one class
being quasilinear, two of these being ``essentially nonlinear".

We recall that the KdV hierarchy consists of the symmetries of
the third order KdV equation at every odd order. The KdV hierachy is characterized by the existence of conserved densities that are
quadratic in the highest derivative at any order. At the fifth order, there
are two basic hierarchies that start at this order, i.e., that
are not symmetries of third order equations.  These are the
Sawada-Kotera and Kaup hierarchies that are derived from a
third order Lax operator. Their symmetries give integrable
equations at odd orders that are not divisible by $3$.
Similarly, every third order conserved density is trivial.  The
equations  that are related to these three hierarchies via
Miura type transformations may have quite different appearance
and form a large list.

In the following  years, the search for new hierarchies of
integrable equations starting at higher orders turned out to be
fruitless; the situation was clarified by  Wang and Sanders,
who proved that scale homogeneous scalar integrable evolution
equations of orders greater than or equal to seven are
symmetries of lower order equations \cite{SW98}.  In subsequent
papers these results were extended to the cases where negative
powers are involved \cite{SW2000} but the case where $F$ is
arbitrary remained open.

The general case where the functional form of $F$ is arbitrary
was studied in the references \cite{B2005} and  \cite {MB2009}.
The first result in this direction has been  obtained in
\cite{B2005}, where the canonical densities $\rho^{(i)}$,
$i=1,2,3$ were computed for evolution equations of arbitrary
order $m$. It was first proved that, up to total derivatives,
higher order $(m \geq 7)$ conserved densities are at most
quadratic in the highest derivative.  Then assuming that an
evolution equation $u_t=F(x,t,u,\dots,u_m)$ admits a conserved
density $\rho^{(1)}=Pu_{m+1}^2+Qu_{m+1}+R$, where $P$, $Q$, $R$
are functions independent of $u_{m+1}$ it has been  shown that
for $m\ge 7$, $PF_{mm}=0$  \cite{B2005}. Finally, it was shown
that the coefficient $P$ in  the canonical density
$\rho^{(1)}$, has the form $P=F_m$ \cite{B2005}, hence it was
concluded that evolution equations of order $m\ge 7$ that admit
the canonical density $\rho^{(1)}$ are quasilinear. In the
proofs of these results, the  remarkable was the fact that the
explicit form of the conserved densities was needed only at the
last stage, to prove that $PF_{mm}=0$ implies $F_{mm}=0$.
Furthermore the derivations used the dependency of $F$ and $P$
on $u_m$ only and the functions $Q$ and $R$ never appeared in
the computations.

In  \cite{MB2009} the same scheme was applied to the
quasilinear equations and it was proved that if the canonical
densities $\rho^{(i)}$, $i=1,2,3$ are conserved then the
evolution equation has to be polynomial in the derivatives
$u_{m-1}$ and $u_{m-2}$.  In the derivation of these results,
in many places we had first assumed the existence of a
``generic" conserved density of a specific form to obtain a
polynomiality result, then we have shown that there is in fact
a canonical density of the required form.  In this work also,
remarkably, all polynomiality results involved the dependencies
of the unknown functions on the top order derivatives, i.e., on
$u_{m-1}$, if
$u_t=A(x,t,u,\dots,u_{m-1})u_m+B(x,t,u,\dots,u_{m-1})$ and so
on.

These observations above lead to the definition of a graded
algebra structure \cite{M2008} which will be the main subject
of this paper.

The classification of $5$th order, constant separant  evolution
equations is given in \cite{MSS91}. The non-constant separant
case is studied by the MSS method in \cite{G2012} where
``KdV-like" equations are defined to be the ones that admit an
unbroken sequence of conserved densities at all orders.
Although the classification is not complete, it has been shown
that the non-constant separant KdV-type equations are  of the
form $u_t=a^5 u_5+B u_4^2 +C u_4 +G$, where $B$, $C$ and $G$
are polynomial in $a$ and $u_3$ and a new  exact solution is
given \cite{G2012}. This class of equations as well as the ones
that we present in Section 4, are expected to belong to the
hierarchy of essentially nonlinear equations of the third
order, given by Eqn.(3.3.9) of \cite{MSS91}.

Recall that eventhough all quasilinear third order equations
are shown to be either linearizable or transformable to the
Korteweg-deVries (KdV)  equation, the Krichever-Novikov
equation is possibly an exception
\cite{HSS95},\cite{HS2005},\cite{LWY2011}.

In the present paper, we shall first introduce the grading
scheme mentioned above and prove its main properties,
namely its invariance under integrations by parts and the
``dependency of the top level on the top derivative".
We shall give the top level parts of the candidates for integrable equations at
orders $m=7,9,11,13,15$. An explicit form for integrable evolution equations of order $m=7,9$ and a closed form for orders $m=11,13,15$ where the explicit form of the coefficient $B$ is given.

The notation and terminology is reviewed in Section 2 and the
level grading is introduced in Section 3. In Section 4 we give
the applications of level homogeneity to the classification of
evolution equations of order $m\geq7.$ Results and discussions
are given in Section 5.

\section{ Notation and terminology}
\subsection{Notation}

Let $u=u(x,t)$. A function $\varphi $ of  $x$, $t$, $u$ and the
derivatives of $u$ up to a fixed but finite order, denoted by
$\varphi[u],$ will be called a ``differential function''
\cite{O93}.
 We shall assume that $\varphi$ has partial
derivatives of all orders. For notational convenience, we shall
denote indices by subscripts or superscripts in parenthesis
such as in $\alpha_{(i)}$ or $\rho^{(i)}$ and reserve
subscripts without parentheses for partial derivatives, i.e.,
for $u=u(x,t)$,
$$ u_0=u, \quad u_t={\partial u \over \partial t},\quad
 u_x={\partial u\over \partial x},\quad
 u_k={\partial^k u\over\partial x^k}$$
and for $ \varphi=\varphi(x,t,u,u_1,\dots,u_n)$,
$$     \varphi_t={\partial \varphi \over \partial t},\quad
\varphi_x={\partial \varphi\over \partial x},\quad
\varphi_k={\partial \varphi\over \partial u_k}.$$
  If $\varphi$ is a differential function, the total derivative with
respect to $x$ is denoted by $D\varphi$ and it is given by
\bea D\varphi=\sum_{i=0}^n \varphi_i  u_{i+1}  + \varphi_x. \eea
Higher order derivatives can be computed by applying the
binomial formula as given below,
\bea D^{k}\varphi= \sum_{i=0}^n \left[\sum_{j=0}^{k-1} {k-1
\choose j}\left( D^{j}\varphi_i\right) u_{i+k-j}\right] +
D^{k-1}\varphi_x. \label{turev} \eea If $u_t=F[u]$, then the
total derivative of $\varphi$ with respect to $t$ is given by
\bea D_t\varphi=\sum_{i=0}^n \varphi_i  D^iF  + \varphi_t.
\label{reel} \eea

The ``order" of a differential  function $\varphi[u]$, denoted by $ord(\varphi)=n$
is the order of the highest derivative of $u$ present in
$\varphi[u]$. The total derivative with respect to $x$
increases the order by one. From the expression of the total
derivative with respect to $t$ given by $(\ref{reel})$ it can
be seen that if $u$ satisfies an evolution equation of order
$m,$ $D_t$ increases the order by $m$.

Equalities up to total derivatives with respect to $x$ will be
denoted by $\cong$, i.e.,
$$ \varphi\cong \psi \ \ \ {\rm if \ and \ only \ if } \ \ \
\varphi=\psi+D\eta. $$

The effect of the integration by parts on monomials is
described as follows. Note that if a monomial is non-linear in
its highest derivative we cannot integrate by parts and reduce
the order. Let $k<p_1<p_2<\dots<p_l<s-1$ and $\varphi$ be a
function of $x,t,u,u_1,\dots,u_k$. Then \bea \varphi
u_{p_1}^{a_1}\dots u_{p_l}^{a_l}u_s&\cong& -D\left(\varphi
u_{p_1}^{a_1}\dots u_{p_l}^{a_l}\right) u_{s-1}, \nonumber\cr
\varphi u_{p_1}^{a_1}\dots u_{p_l}^{a_l}u_{s-1}^pu_s&\cong&
-\textstyle\frac{1}{p+1} D\left(\varphi u_{p_1}^{a_1}\dots
u_{p_l}^{a_l}\right) u_{s-1}^{p+1}.\nonumber\cr \eea The
integrations by parts are repeated successively until one
encounters a ``non-integrable monomial" of the following form:
$$u_{p_1}^{a_1}\dots u_{p_l}^{a_l}u_{s}^p,\quad p>1.$$
The order of a differential monomial is not invariant under
integration by parts, but we will show in the next section that
its level decreases by one under integration by parts
\cite{M2008}. This will be the rationale and the main advantage
of using the level grading.
% Section 3
%
\section { The Ring of Polynomials and
``level-grading"}

The scaling symmetry and scale homogeneity are well known properties
of polynomial integrable equations.
We recall that scaling symmetry is the invariance of an equation under the transformation
$u\to \lambda^{a}u$, $x\to \lambda^{-1} x$, $t\to \lambda^{-b} t$.
If $a=0$, then scale invariant quantities may be non-polynomial
and the scaling weight is just the order of differentiation. In
the early stages of our investigations we have noticed that if
$F$ is a function of the derivatives of $u$ up to order say
$k$, and we differentiate $F$ say $j$ times, the resulting
expressions are polynomial in $u_{k+1},\dots, u_{k+j}$. Furthermore,
 the sum of the order of differentiations   exceeding  $k$ has some type of invariance.
This remark led us to the definition of ``level grading" as a
generalization of the scaling symmetry for the case $a=0$, as
a graded algebra structure.

We consider the ring of functions of $x,t,u,\dots ,u_k$ the modules generated by the derivatives $u_{k+1},\dots$.
This set up is given a graded algebra structure as described below.

Let $M$ be an algebra over a ring $K$. If we can write $M=\oplus_{i\in N}M_i$, as a direct sum of its submodules $M_i$, with the property that
$M_iM_j\subseteq M_{i+j}$, then we have a ``graded algebra" structure on $M$.  For example if $K=R$ and $M$ is the algebra of  polynomials
in $x$ and $y$, then the $M_i$'s may be chosen as the submodule consisting of homogeneous polynomials of degree $i$.
In the same example, we may also consider the submodules consisting of polynomials of degree $i$ (not necessarily homogeneous)
that we denote by $M^i$. Then $M^i$ is the direct sum of the submodules $M_j$, $j$ ranging from zero to $i$.
It follows that the full algebra $M$ can be written as a sum of the submodules $M_i$, but the sum is no more direct.
This structure is called a ``filtered algebra".  The formal definitions are given below.

\vskip 0.2cm \noindent \textbf{Definition 3.1}: {Let $K$ be a
ring and $M$ be an algebra over $K$ and $M_i$ be submodules of
$M$. The decomposition of $M$ to a direct sum of submodules: \
$M=\oplus_{i\in N}M_i$ \ and \ $M_iM_j\subseteq M_{i+j}$ is
called a graded algebra structure on $M.$ Given a graded
algebra $M$, we can obtain an associated ``filtered algebra"
$\tilde{M}=M,$ \cite{V74}, by defining $M=\sum_{i\in
N}\tilde{M}_i$ where $\tilde{M}_i=\oplus_{j=0}^i M_j.$ } \hfill
$\Box$ \vskip 0.1cm

If the algebra $M$ is characterized by a set of generators, then the submodules $M_i$ can also be characterized similarly.
If $K^{(k)}$ be the ring of $C^{\infty}$ functions of
$x,t,u,\dots,u_k$, and $M^{(k)}$ is the polynomial algebra over
$K^{(k)}$ generated by the set
$$S^{(k)}=\{u_{k+1},u_{k+2},\dots\},$$
a monomial in $M^{(k)}$ is a product of a finite number of
elements of $S^{(k)}$. We define the ``level above $k$"  of a
monomial as follows:
\vskip 0.2cm
\noindent \textbf{Definition 3.2}:  Let
$\mu=u_{k+j_1}^{a_1}u_{k+j_2}^{a_2}\dots u_{k+j_n}^{a_n}$ be a
monomial in $M^{(k)}$. The {\bf level of $\mu$ above $k$}, is
defined by
$$lev_k(\mu)=a_1j_1+a_2j_2+\dots+a_nj_n.$$
The level of the differential operator $D$ is defined to be $1$.
 The level of a
pseudo-differential operator is thus
  $lev_k(\varphi D^j) = lev_k(\varphi) + j$.\hfill $\Box$
\vskip 0.2cm

It can be seen that for any two monomials $\mu$ and
$\tilde{\mu}$,
$$ lev_k(\mu\tilde{\mu})=lev_k(\mu)+lev_k(\tilde{\mu}). $$
The ``level above $k$" gives a graded algebra
structure to $M^{(k)}$. Monomials of a fixed level $p$ form a
free module $M^{(k)}_p$ over $K^{(k)}$ and we denote its set of
generators by $S_p^{(k)}$.
If $a$ is a polynomial in $\tilde{M}_p^{(k)}$, \ $a=\sum_{p\geq
0}a_p$ where $a_p \in {M}_p^{(k)}$, \ $a_p$ is called the
homogeneous component of $a$ of level $p$ above $k$.
\vskip 0.2cm
\noindent \textbf{Definition 3.3} Let $a$ be a polynomial in
$\tilde{M}_p^{(k)}$.  The image of a polynomial $a$ under the
natural projection  $$\pi: \tilde{M}_p^{(k)} \to {M}_p^{(k)}$$
 denoted by $\pi(a)$ is called the ``top level part of $a$".
\hfill $\Box$
\vskip 0.2cm

We will now present certain results that demonstrate the
importance of the level grading.  These will be the proofs that
partial derivatives with respect to $u_i$, total derivatives
with respect to $x$, total derivatives with respect to $t$, the
integration by parts hence the conserved density conditions are
filtered algebra maps.

\vskip 0.2cm
\noindent \textbf{ Proposition 3.4} { \it If $\varphi$ is
level homogeneous of level $p$ above $k$  and if
$\frac{\partial{\varphi}}{\partial{u_{k+j}}}\neq 0 $, then the
partial derivative of $\varphi $ with respect to $u_{k+j}$
has level $p-j$, for $ j\geq 0$.}

\vskip 0.2cm \noindent \textbf{Proof:} Let $\varphi \in
M_p^{(k)}$ and $\varphi$ be level homogeneous above $k$ of
level $p$.  Then $\varphi$ is a linear combination of monomials
of level $p$;
$M_p^{(k)}=\left<u_{k+p},u_{k+p-1}u_{k+1},u_{k+p-2}u_{k+2},u_{k+p-2}u_{k+1}^2,\dots
\right>$.  Clearly if
$\frac{\partial{\varphi}}{\partial{u_{k+j}}}\neq 0 $, the
effect of differentiation with respect to
$\frac{\partial{\varphi}}{\partial{u_{k+j}}} $  decreases the
level by $j$.

\hfill$\Box$

 \vskip 0.2cm
\noindent \textbf{ Proposition  3.5} {\it The total
derivative with respect to $x$, $D$  is a filtered algebra
map $\tilde{M}_p^{(k)}\to \tilde{M}_{p+1}^{(k)}$.}

\vskip 0.1cm \noindent \textbf{ Proof:} Clearly it is
sufficient to consider the effect of $D$ of a product of a
function $\varphi$  in $K^{(k)}$ and a monomial of level $p$
above $k$.  The effect of $D$ on a monomial increases the level
by $1$.  On the other hand $D\varphi= \frac{\partial
\varphi}{\partial u_k}u_{k+1}+\dots$.  In particular, the level
$p+1$ part depend only on $\varphi$ and its derivative with
respect to $u_k$. It follows that $D$ is a filtered algebra
map. \hfill$\Box$

\vskip 0.2cm We now study the effect of integration by parts.
The subset of the generating set $S^{(k)}_p$ of the  module
$M^{(k)}_p,$ consisting of the monomials that are nonlinear in
the  highest derivative and the submodule that it generates are
denoted  respectively by $\bar{S}_p^{(k)}$ and
$\bar{M}^{(k)}_p$.  If a monomial is nonlinear in its highest
derivative it cannot be integrated.  If it is linear, one can
proceed with the integrations by parts until a term that is
nonlinear in its highest derivative is encountered.  By virtue
of the propositions above, these operations will be  filtered
algebra maps.

 \vskip 0.2cm
\noindent \textbf{ Proposition  3.6} {\it Let $\alpha$ be a
polynomial in $\tilde{M}^{(k)}_p$.  Then
$$\int \ \alpha= \beta -\int \ \gamma$$
where  $\beta$ belongs to $\tilde{M}^{(k)}_{p-1}$ and $\gamma$
belongs to $\tilde{M}^{(k)}_p$.}
\vskip 0.1cm
\noindent \textbf{ Proof:}
Let $$ \mu=u_{k+i_1}^{a_1}u_{k+i_2}^{a_2} \dots u_{k+i_j}^{a_j},
\quad \quad i_1>i_2> \dots >i_j \quad \quad
i_1a_1+i_2a_2+\dots+i_ja_j=p $$
We have the following three mutually exclusive cases.
\begin{description}
\item [i.] When $ a_1>1$, the monomial is not a total derivative and $\mu\in \bar{S}_{p,n}^{(k)}$. We cannot proceed with integration by parts.
\item [ii.] When $a=1$ the term $\varphi \mu$ where $\varphi\in K^{(k)}$  can be integrated.  For $i_2 < i_1-1$
      $$ \int \varphi \mu =\int
      \varphi u_{k+i_1}^{a_1}u_{k+i_2}^{a_2} \dots u_{k+i_j}^{a_j}=
      \varphi u_{k+i_1-1}^{a_1} u_{k+i_2}^{a_2} \dots u_{k+i_j}^{a_j}-
      \int u_{k+i_1-1}^{a_1} D\left(\varphi u_{k+i_2}^{a_2} \dots u_{k+i_j}^{a_j}\right).$$
\item [iii.] When $a=1$ but $i_2=i_1-1$ then
$$ \int \varphi \mu =\int
      \varphi u_{k+i_1}^{a_1}u_{k+i_1-1}^{a_2}u_{k+i_3}^{a_3} \dots u_{k+i_j}^{a_j}=
      \frac{u_{k+i_1-1}^{a_2+1}}{a_2+1} u_{k+i_3}^{a_3} \dots u_{k+i_j}^{a_j}-
      \int  \frac{u_{k+i_1-1}^{a_2+1}}{a_2+1}
       D\left(\varphi u_{k+i_3}^{a_3} \dots u_{k+i_j}^{a_j}\right).$$
\end{description}
In (i) and (ii), the level of the term that has been integrated decreases by $1$ while the terms under the integral sign have levels $p$ or lower.
\hfill$\Box$
\vskip 0.2cm

We will now give an example that illustrates the effect of
total derivatives and integration by parts.
\vskip 0.2cm
\noindent \textbf{Example 3.7} Let
$$ R = \varphi u_8 +
\psi u_7u_6 + \eta u_6^3,$$
 where $\varphi, \psi, \eta \in
K^{(5)}$ be a polynomial in $M^{(5)}_3$. It can easily be seen
that $DR$ is a sum of polynomials in $M_4^{(5)}$ and
$M_3^{(5)}$

\bea DR &=& \underbrace{\varphi u_9+(\varphi_5+\psi)u_8u_6+\psi
u_7^2+(\psi_5+3\eta)u_7u_6^2+\eta_5u_6^4}\nonumber \\
& & \quad \quad \quad \quad \quad  \quad \quad \quad M_4^{(5)} \nonumber \\
&+&\underbrace{(\varphi_4
u_5+\dots+\varphi_x)u_8+(\psi_4 u_5+\dots+\psi_x)u_7u_6+(\eta_4 u_5+\dots + \eta_x)u_6^3} \nonumber \\
& & \quad \quad \quad  \quad  \quad  \quad  \quad  \quad \quad
M_3^{(5)} \nonumber\eea
Note  that the projection to
$M^{(5)}_4$ depends only on the derivatives with respect to
$u_5$. We later prove that this holds in general.

For convenience, we define the operator $D_0$ to denote the
part of $D\phi $ on lower order derivatives by,
$D\varphi=\varphi_k u_{k+1}+ D_0\varphi$ as a sum of level 1
and level 0 terms. It follows that $D^2 \varphi=\varphi_k
u_{k+2} +\varphi_{kk} u_{k+1}^2 +(D_0\varphi)_k u_{k+1} +
D_0(D_0 \varphi)$ is a sum of level 2, level 1 and level 0
terms. The integration by parts of $R$ gives: \bea
\displaystyle  \int
R dx= & & \underbrace{ \varphi u_7  + \frac{1}{2}[\psi -\varphi_5]u_6^2 - D_0 \varphi u_6} \nonumber \\
& &  \quad \quad \quad \quad \quad \quad \quad M_2^{(5)} \nonumber \\
& & +\int \underbrace{\left[
\left[\frac{1}{2}\varphi_{55}-\frac{1}{2}\psi_5-\eta\right]
u_6^3 +
\left[\frac{1}{2}D_0\varphi_5-\frac{1}{2}D_0\psi+(D_0\varphi)_5\right]u_6^2 + D_0(D_0\varphi)u_6\right]}dx. \nonumber \\
& & \quad \quad \quad \quad \quad \quad \quad \quad \quad \quad
\quad   M_3^{(5)} \nonumber\eea \hfill$\Box$
\vskip 0.2cm
Now we deal with time derivatives.  Given
$u_t=F(x,t,u,...,u_m)$ where $F$ is of order $m$, if
$\rho=\rho(x,t,u,...,u_n)$ is a differential polynomial of order
$n$, then clearly, $D_t\rho$ is of order $n+m$. A similar result holds for level grading.
\vskip 0.2cm \noindent \textbf{ Proposition  3.8} {\it Let
$u_t=F[u]$, where $F$ is a differential polynomial of order $m$
and of level $q$ above the base level $k$. Then $D_t$  is a
filtered algebra map $\tilde{M}_p^{(k)}\to
\tilde{M}_{p+q+k}^{(k)}$.}
\vskip 0.1cm
\noindent \textbf{Proof:} Let $\rho$ be a differential polynomial of order $n$ and of level $p$ above the base level $k$. Then
$$D_t\rho=\rho_t+\sum_{i=0}^k \frac{\partial\rho}{\partial u_i}D^iF+
  \sum_{j=1}^{n-k} \frac{\partial\rho}{\partial u_{k+j}}D^{k+j}F.$$
Note that $\rho_t$ has level at most $p$. Similarly, the  level
of $\frac{\partial\rho}{\partial u_i}$ for $i\le k$, is at most
$p$ hence each of the terms in the first summation are of
levels at most $p+q+i,$ and the sum has level at most $p+q+k.$
In the second summation, $\frac{\partial\rho}{\partial
u_{k+j}}$ has level $p-j$, hence the level of
$\frac{\partial\rho}{\partial u_{k+j}}D^{k+j}F$ is
$(p-j)+(k+j)+q=p+q+k$.
\hfill$\Box$
\vskip 0.2cm

\noindent \textbf{Corollary 3.9} If $F$ is quasi-linear, and
$m=k+q,$ then $D_t$ increases the level by $m.$\hfill$\Box$

We will now prove a very useful proposition stating that the
top level depends only on the dependency of the coefficients on
$u_k$.
\vskip 0.2cm
\noindent \textbf{ Proposition  3.10} {\it Let $\rho$ be a
differential polynomial in  $\tilde{M}^{(k)}_p$. Then the
projection $\pi (D^j \rho)$ depends only on the dependency of
the coefficients in $\rho $ on $u_k$.}
\vskip 0.1cm
\noindent {\bf Proof.} Let $\rho=\sum_i \varphi_i P_i$ where
$\varphi_i \in K^{(k)}$ and $P_i \in M_p^{(k)}.$ Without lost
of generality $\rho=\varphi P$ and $ P=u_{i_1}^{a_1} \dots
u_{i_n}^{a_n}.$ Here $lev(\rho)=p.$

\bea D\rho &=& D\varphi P+\varphi D P = \left[\varphi_x +
\sum_{i=0}^{k-1} \frac{\partial{\varphi}}{\partial{u_i}}u_{i+1}
+ \frac{\partial{\varphi}}{\partial{u_k}}u_{k+1}\right]P +
\varphi D P \nonumber \\
&=& \underbrace{\left[\varphi_x + \sum_{i=0}^{k-1}
\frac{\partial{\varphi}}{\partial{u_i}}u_{i+1}  \right]P} +
\underbrace{\frac{\partial{\varphi}}{\partial{u_k}}u_{k+1}P +
\varphi D P}
\nonumber \\
& & \quad \quad \quad \quad  M_p^{(k)} \quad \quad \quad \quad
\quad \quad \quad \quad M_{p+1}^{(k)}\eea

It follows that the projection $\pi(D^j\rho)$ is independent of
$ \displaystyle \frac{\partial{\varphi}}{\partial{}u_j}$ for
$j<k$ and independent of $ \displaystyle
\frac{\partial{\varphi}}{\partial{u_x}}$
\hfill$\Box$.
\vskip 0.2cm

It follows that in the conserved density computations, if
$\rho$ and $F[u]$ are level homogeneous, then $\rho_t$ up to
total derivatives is also level homogeneous. This is a very
important and useful result.

\section{Application of ``level homogeneity" structure on the classification
problem}

In this section we apply the ``level homogeneity" structure to
the classification of scalar evolution equations of orders
$m\geq 7.$ In \cite{MB2009} we have shown that if $F$ is
integrable in the sense of admitting a formal symmetry,  then
it is of the form
$$ F=u_t= a^{m}u_{m} + B u_{m-1}u_{m-2} + \dots $$
where $a$, $B$, $C$, $G$, $H$ and  $K$ are functions of $x$, $t$, $u$, $u_i$,
$i\le m-3$,
i.e., it is level homogeneous above level
$m-3$.

Here we start from this form of $F$ which is level
homogeneous above level $m-3$.  We shall assume that the conserved densities
$\rho^{(-1)}$, $\rho^{(1)}$ and  $\rho^{(3)}$ are non-trivial.
  The cases where either of these are trivial will be dealt with elsewhere.  We characterize such equations as ```KdV-like". It is well known that the canonical densities of even order are trivial, hence we give the definition as below.
\vskip 0.2cm
\noindent \textbf{ Definition 4.1}
An evolution equation $u_t=F[u]$ is called ``KdV-like" if its sequence of odd numbered canonical densities is nontrivial.\hfill$\Box$
\vskip 0.2cm

When we substitute the form of $F$ given above in the canonical
conserved densities $\rho_c^{(i)}$ and we integrate by parts we
can see that the canonical densities are of the form  given
below and we use the subscript $c$ to denote canonical
quantities. Alternatively,  if haven't the explicit expression
of the canonical densities we could assume that the evolution
equation admits an infinite sequence of level homogeneous
conserved densities that we call ``generic conserved
densities". In fact, from a computational point of view, it is
preferable  to use  the generic conserved densities and compare
with the explicit form of the canonical densities whenever
necessary. The generic quantities are labeled by the excess of
the order of the highest derivative above the base level. If
$k$ be the base level, $m$ the order and $F_m=A=a^m$, then the
generic form of the conserved densities are as follows: \bea
\rho_c^{(-1)}=\rho^{(0)}&=&A^{-1/m}=a^{-1} \nonumber \\
\pi(\rho_c^{(1)})\cong \rho^{(1)}&=& P^{(1)} u_{k+1}^2 \nonumber \\
\pi(\rho_c^{(3)})\cong\rho^{(2)}&=& P^{(2)} u_{k+2}^2 + Q^{(2)}
u_{k+1}^4. \label{yogun}\eea

\noindent \textbf{Remark 4.2}
Recall that conserved densities can be given up to total derivatives. Thus a generic conserved density of of order $k+j$ is a polynomial in the monomials $M_{2j}^{(k)}$, as given in Appendix B.\hfill$\Box$
\vskip 0.2cm

We will outline below the steps leading to the classification
of the top level parts of the integrable equations of odd
orders $m=7,\dots, 15$ for scalar evolution equations admitting
the (nontrivial) canonical conserved densities $ \rho_c^{(1)}$,
$\rho_c^{(2)}$, $\rho_c^{(3)}$. In particular the nontriviality
of $\rho_c^{(3)}$ will be crucial. \vskip 0.2cm

\noindent {\bf Step 1. $k=m-3$, $m=7,9,11,13,15.$} We begin our
computations for $k=m-3$, for $m=7,9,11,13,15$.  In
\cite{MB2009}it has been shown that any $m$ integrable
evolution equations are of the form
 \be u_t=F=a^{k+3}u_{k+3} + B u_{k+2}u_{k+1} + C u_{k+1}^3,\quad \quad  m \geq 7. \label{bir}
\ee For each order, we compute the conserved density
conditions, integrate by parts and collect the top level terms.
The solutions of these equations give that all the coefficients
$B,$ \ \ $C$ are functions of $a$ and the derivatives of $a$
with respect to $u_k$ of various orders and $a$ is independent
of $u_k.$  This implies that $F$ is level homogeneous over
$K^{(m-4)}$. \vskip 0.2cm

\noindent {\bf Step 2. $k=m-4$, $m=7,9,11,13,15.$} For $k=m-4$
the generic form of the evolution equation is: \be F=u_t=
a^{k+4}u_{k+4} + B u_{k+3}u_{k+1} + C u_{k+2}^2 + G u_{k+2}
u_{k+1}^2 + H u_{k+1}^4, \ m \geq 7, \label{iki} \ee where the
coefficients depend on $u_i$ for $i\le m-4$. The top level
parts of the conserved densities have the same form.  Computing
the conserved density conditions and integrating by parts we
obtain systems of equations.  For $m=7$, we find that $a$
satisfies the third order differential  equation
 $$ \displaystyle a_{333}-9 a_{33} a_3 a^{-1} +12a_3^3a^{-2} \label{ayni}$$
 and the classification of the $7$th order equations is not pursued further.
 For $m>7$, we have $a_{m-4}=0$ and it turns out that $F$ is level homogeneous over  $K^{(m-5)}$.
\vskip 0.2cm

\noindent {\bf Step 3. $k=m-5$, $m=9,11,13,15.$} For $k=m-5$
the generic form of the evolution equation is: \be
F=u_t=a^{k+5}u_{k+5} + B u_{k+4} u_{k+1} + C u_{k+3} u_{k+2} +
E u_{k+3}u_{k+1}^2 + G u_{k+2}^2 u_{k+1} + H u_{k+2} u_{k+1}^3
+ K u_{k+1}^5, \ \ m \geq 9 \label{uc} \ee The conserved
density conditions imply that $F$ is level homogeneous over
$K^{(m-6)}$, for $m=9,11,13,15.$

\vskip 0.2cm \noindent {\bf Step 4. $k=m-6$, $m=9,11,13,15.$}
For $k=m-6$ the generic form of the evolution equation is: \bea
F&=&a^{k+6}u_{k+6} + B u_{k+5}u_{k+1} + C u_{k+4}u_{k+2} + E
u_{k+4}u_{k+1}^2 + G u_{k+3}^2 + H
u_{k+3}u_{k+2}u_{k+1}\nonumber \\
& & + K u_{k+3}u_{k+1}^3 + L u_{k+2}^3 + M u_{k+2}^2u_{k+1}^2 +
N u_{k+2}u_{k+1}^4 + P u_{k+1}^6, \ \ m \geq 9 \label{dort}
\eea We note that the expression of $F$ given above is a linear
combination of the monomials in the generating set of
$M_6^{(k)}$, as given in Appendix A. For $m=9$, surprisingly we
find that $a$ satisfies the same equation as above
$(\ref{ayni})$, and for $m>9$ we find that $a_{m-6}=0$, and it
follows that $F$ is level homogeneous over $K^{(m-7)}.$

\vskip 0.2cm \noindent {\bf Step 5. $k=m-7$, $m=11,13,15.$} At
this step, $F$  is a linear combination of the monomials in the
generating set of $M_7^{(k)}$, given in Appendix A and we omit
the explicit expression here. The conserved density conditions
imply that $F$ is level homogeneous over $K^{(m-8)}.$

\vskip 0.2cm \noindent {\bf Step 6. $k=m-8$, $m=11,13,15.$} $F$
is now a linear combination of the monomials in the generating
set of $M_8^{(k)}$, given in Appendix A.  The conserved density
conditions imply that for $m=11$, $a$ satisfies the equation
above $(\ref{ayni})$ and for $m>11$, $a_{m-8}=0$. It follows
that for $m>11$, $F$ is level homogeneous over $K^{(m-9)}.$

\vskip 0.2cm \noindent {\bf Step 7. $k=m-9$, $m=13,15.$} $F$ is
a linear combination of the monomials in the generating set of
$M_9^{(k)}$, given in Appendix A.  The conserved density
conditions imply that $a_{m-9}=0$ and $F$ is level homogeneous
over $K^{(m-10)}.$

\vskip 0.2cm \noindent {\bf Step 8. $k=m-10$, $m=13,15.$} $F$
is  a linear combination of the monomials in the generating set
of $M_{10}^{(k)}$, given in Appendix A.  For $m=13$,  $a$
satisfies that equation above $(\ref{ayni})$ while for $m>11$,
$a_{m-10}=0$ and $F$ is level homogeneous over $K^{(m-11)}.$

\vskip 0.2cm \noindent {\bf Step 9. $k=m-11$, $m=15.$} $F$  is
a linear combination of the monomials in the generating set of
$M_{11}^{(k)}$, given in Appendix A.  The conserved density
conditions imply that $a_{m-11}=0$ and $F$ is level homogeneous
over $K^{(m-12)}.$

\vskip 0.2cm \noindent {\bf Step 10. $k=m-12$, $m=15.$} $F$  is
a linear combination of the monomials in the generating set of
$M_{12}^{(k)}$, given in Appendix A.  The conserved density
conditions imply that $a$ satisfies the equation above
$(\ref{ayni})$. \vskip 0.2cm

It is a remarkable fact that at all orders $m\geq 7$,  the
separant $a$ satisfies the following equation.

\be \displaystyle a_{333}-9 a_{33} a_3 a^{-1} +12a_3^3a^{-2}=0
\label{tanimlar1}.\ee Using the substitution $a=Z^{-1/2}$ in
the equation above, we obtain $Z_{333}=0$, hence \be
a=\left(\alpha u_3^2 +\beta u_3+\gamma\right)^{-1/2}
\label{tanimlar2},\ee where $\alpha$, $\beta$, $\gamma$ are
functions of $x$, $t$, $u$, $u_1$ and $u_2$ in general. Here as
we are interested in the top level form of the equations, the
dependencies on these derivatives are irrelevant.
 We give below certain expressions that are useful for a controlled substitution in the conserved density conditions.

\subsection{Classification of scalar evolution equations of order $m=7$}
\noindent Generally for coefficients that depend on lower
orders we will use capital letters. We summarize this
computations in two parts. First our base level is $m-3=4.$ We
work with scalar evolution equations of order $m=7,$ \be
u_t=a^7 u_7+B u_5u_6+C u_5^3\ee and the generic conserved
densities $\rho^{(0)}, \rho^{(1)}, \rho^{(2)}$ given in
$(\ref{yogun})$ where the coefficients $a,B,C$ and $ P^{(1)},
P^{(2)},Q^{(2)}$ depend on
  $u_4.$\\
  We get that all the coefficients $B,C,P^{(1)},...$ are functions of $a,$ and the derivatives of $a$ with respect to $u_4$ of various orders. Finally we get
  that the derivative of $a$ with respect to $u_4$ is zero, $$a_4=0,$$ which implies that all the coefficients vanishes. Then we reduce the base level $u_4$ by one.\\
\noindent In the second part our base level is $m-4=3.$ We work
with scalar evolution equations of order $m=7,$ \be u_t=a^7
u_7+B u_6u_4+C u_5^2+G u_5u_4^2+H u_4^4 \ee and the generic
conserved densities $\rho^{(0)},\rho^{(1)},\rho^{(2)}$ given in
$(\ref{yogun})$ where the coefficients \\ $a,B,C,G,H$ and $
P^{(1)},P^{(2)},Q^{(2)}$ depend on $u_3.$

In this step we obtain, $P^{(1)}=P^{(10)} a^5,$ \  \
$P^{(2)}=P^{(20)} a^7$ \ and
\bea \displaystyle u_t&=&a^7 u_7 + 14a_3 a^6 u_6u_4 + \frac{21}{2}a_3 a^6 u_5^2 + a^5\left(\frac{35}{2}a_{33} a + 63a_3^2\right)u_5 u_4^2 \nonumber \\
&+&  a^4 a_3\left( \frac{399}{8}a_{33}
a-\frac{21}{4}a_3^2\right)u_4^4. \eea

When the conserved densities are:
\bea \rho^{(1)}&=&P^{(10)}a^5 u_4^2 \nonumber \\
\rho^{(2)}&=&P^{(20)} a^7u_5^2 + P^{(20)} a^5
\left(-\frac{7}{4}a_{33} a-\frac{7}{2}a_3^2\right)u_4^4
\label{conserved}\eea \vskip 1.5cm

Thus seventh order integrable scalar evolution equations have
the following form:

\bea \displaystyle u_t&=&a^7 u_7 - 7 a^9 z_3 u_4 u_6
-\frac{21}{4} a^9 z_3 u_5^2 \nonumber \\
& & + \left(-\frac{231}{8} a^2 p + 98 \alpha \right)a^9 u_4^2
u_5 \nonumber \\
& & + \left( \frac{1155}{64} a^2 p - \frac{189}{4} \alpha
\right)a^{11} z_3 u_4^4 .\eea

Where $a$ satisfies the equations given in $(\ref{tanimlar1}),$
$(\ref{tanimlar2}).$

\subsection{Classification of scalar evolution equations of order $m=9$}
In this section we compute $9th$ order evolution equations in
$4$ steps. First we work with \be u_t=a^9u_9+B u_7u_8+ C u_7^3
\ee and the generic conserved densities
$\rho^{(0)},\rho^{(1)},\rho^{(2)}$ given in $(\ref{yogun})$
where the coefficients $a,B,C,$ and $P^{(1)},P^{(2)},Q^{(2)}$
depend on $u_6.$

We get that all the coefficients $B,C,P^{(1)},...$ are
functions of $a$ and the derivatives of $a$ with respect to
$u_6$ of various orders. Finally we get that the derivative of
$a$ with respect to $u_6$ is zero, $$a_6=0,$$ which means that
all the coefficients vanishes. Then we reduce the base order
$u_6$ by one.

In the second step we work with scalar evolution equations of
order $m=9,$ \be u_t=a^9 u_9+B u_8u_6+C u_7^2+G u_7u_6^2+H
u_6^4 \ee and the generic conserved densities  $\rho^{(0)},
\rho^{(1)}, \rho^{(2)}$ given in $(\ref{yogun})$ where the
coefficients $a,B,C,G,H$ and $P^{(1)}, P^{(2)},Q^{(2)}$ depend
on $u_5.$

In this step also we get that all the coefficients
$B,C,P^{(1)},...$ are functions of $a,$ and the derivatives of
$a$ with respect to $u_5$ of various orders. Finally we get
that the derivative of $a$ with respect to $u_5$ is zero,
$$a_5=0,$$ which means that all the coefficients vanishes. Then we reduce the base order $u_5$ by one.

In the third step we obtain the similar results where all the
coefficients $B,C,P^{(1)},...$ are functions of $a,$ and the
derivatives of $a$ with respect to $u_4$ of various orders and
$$a_4=0.$$ The scalar evolution equations of order $m=9,$ that we work with, in this step,
is \be u_t=a^9 u_9+B u_8u_5+C u_7u_6+G u_7u_5^2+H
u_6^2u_5+Ku_6u_5^3+Lu_5^5 \ee and the generic conserved
densities $\rho^{(0)}, \rho^{(1)}, \rho^{(2)}$ are given in
$(\ref{yogun})$ where the coefficients $a,B,C,G,H,K,L$ and
$P^{(1)},P^{(2)},Q^{(2)}$ depend on $u_4.$ Since we get that
the derivative of $a$ with respect to $u_4$ is zero, and all
the coefficients vanishes, we reduce the base order $u_4$ by
one.

Scalar evolution equations of order $m=9,$ that we use in the
last step computations have the following form. \be u_t=a^9
u_9+B u_8u_4+C u_7u_5+G u_7u_4^2+H
u_6^2+Ku_6u_5u_4+Lu_6u_4^3+Mu_5^3+Nu_5^2u_4^2+Pu_5u_4^4+Qu_4^6
\ee and the generic conserved densities $\rho^{(0)},
\rho^{(1)}, \rho^{(2)}$ given in $(\ref{yogun})$ where the
coefficients $a,B,C,G,H,K,L,M,N,P,Q$ and
$P^{(1)},P^{(2)},Q^{(2)}$ depend on $u_3.$

In this step we get $P^{(1)}=P^{(10)} a^5,$ \  \
$P^{(2)}=P^{(20)} a^7.$

The conserved densities are the same as order $m=7$ given in
$(\ref{conserved}).$

Thus ninth order integrable scalar evolution equations have the
following form: \bea u_t&=& a^9 u_9 -\frac{27}{2}a^{11} z_3
u_8u_4-\frac{57}{2}a^{11}z_3 u_7u_5
+a^11\left(-\frac{825}{8}a^2 p+360 \alpha \right)u_7 u_4^2
\nonumber\\ & & -\frac{69}{4}a^11 z_3 u_6^2 +
a^11\left(-\frac{1419}{4}a^2
p+1230 \alpha \right)u_6u_5u_4 \nonumber \\
& & +a^13 z_3\left(\frac{2145}{4}a^2 p-1485 \alpha
\right)u_6u_4^3 +a^{11}\left(-\frac{671}{8} a^2 p+290
\alpha\right)u_5^3 \nonumber \\
& & +a^{13}z_3 \left(\frac{35607}{32}a^2 p-\frac{6105}{2}
\alpha \right)u_5^2u_4^2+a^{13} \left(\frac{255255}{128} a^4
p^2-\frac{94809}{8}a^2 \alpha p+16335 \alpha^2\right)u_5u_4^4
\nonumber \\
& & +a^{15} z_3 \left(-\frac{425425}{512}a^4
p^2+\frac{135135}{32}a^2 \alpha p-\frac{19305}{4}
\alpha^2\right)u_4^6.\eea

Where $a$ satisfies the equations given in $(\ref{tanimlar1}),$
$(\ref{tanimlar2}).$

\subsection{Classification of scalar evolution equations of order $ \bf m=11,13,15$ }
 In this section we give the final form of equations of scalar integrable evolution equations of order $m=11,13,15$ that are computed in
 $6,8,10$ steps respectively.

The conserved densities used are the same as order $m=7$ given
in $(\ref{conserved}).$ \ For each equation of order
$m=11,13,15,$ \  $a$ satisfies the equations given in
$(\ref{tanimlar1}),$ $(\ref{tanimlar2}).$

Scalar integrable evolution equations of order $m=11$ has the
form of: \bea u_t &=& a^{11} u_{11}+B_0 u_{10} u_{4} + B_1 u_9
u_5 + B_2 u_9 u_4^2 + B_3 u_8 u_6 + B_4 u_8 u_5 u_4 + B_5 u_8 u_4^3 \nonumber \\
& & + B_6 u_7^2 + B_7 u_7 u_6 u_4 + B_8 u_7 u_5^2+ B_9 u_7 u_5
u_4^2 +B_{10} u_7 u_4^4 + B_{11}u_6^2 u_5 \nonumber \\
& &  + B_{12} u_6^2 u_4^2 +B_{13}u_6u_5^2 u_4 +B_{14} u_6 u_5
u_4^3 + B_{15} u_6 u_4^5 + B_{16} u_5^4 + B_{17} u_5^3 u_4^2
\nonumber \\
& & + B_{18} u_5^2 u_4^4 + B_{19} u_5 u_4^6+ B_{20} u_4^8\eea

where $B_0=-22 a^{13} z_3.$

Scalar integrable evolution equations of order $m=13$ has the
form of: \bea u_t &=& a^{13} u_{13}+B_0 u_{12} u_{4} + B_1
u_{11}u_5 + B_2 u_{11} u_4^2 + B_3 u_{10} u_6 + B_4 u_{10} u_5 u_4 + B_5 u_{10} u_4^3 \nonumber \\
& & + B_6 u_9 u_7 + B_7 u_9 u_6 u_4 + B_8 u_9 u_5^2+ B_9 u_9
u_5 u_4^2 +B_{10} u_9 u_4^4 + B_{11} u_8^2 + B_{12} u_8 u_7 u_4
\nonumber \\
& & +B_{13}u_8u_6 u_5 +B_{14} u_8 u_6 u_4^2 + B_{15} u_8 u_5^2
u_4 + B_{16} u_8 u_5 u_4^3 + B_{17} u_8 u_4^5 + B_{18} u_7^2
u_5
\nonumber \\
& & + B_{19} u_7^2 u_4^2+ B_{20} u_7 u_6^2 + B_{21} u_7 u_6 u_5
u_4 + B_{22} u_7 u_6 u_4^3 + B_{23} u_7 u_5^3 + B_{24} u_7
u_5^2 u_4^2 \nonumber \\
& & + B_{25} u_7 u_5 u_4^4 + B_{26} u_7 u_4^6 + B_{27} u_6^3
u_4 + B_{28}u_6^2 u_5^2+ B_{29} u_6^2 u_5 u_4^2 + B_{30} u_6^2
u_4^4 \nonumber \\
& & +B_{31}u_6 u_5^3 u_4 + B_{32} u_6 u_5^2 u_4^3 + B_{33} u_6
u_5 u_4^5 + B_{34} u_6 u_4^7 + B_{35} u_5^5 + B_{36} u_5^4
u_4^2 \nonumber \\
& & + B_{37} u_5^3 u_4^4 + B_{38} u_5^2 u_4^6 + B_{39} u_5
u_4^8 + B_{40} u_4^{10} \eea

where $B_0=-\frac{65}{2}a^{15} z_3.$

Scalar integrable evolution equations of order $m=15$ has the
form of: \bea u_t &=& a^{15} u_{15}+B_0 u_{14} u_{4} + B_1
u_{13}u_5 + B_2 u_{13} u_4^2 + B_3 u_{12} u_6 + B_4 u_{12} u_5 u_4 + B_5 u_{12} u_4^3 \nonumber \\
& & + B_6 u_{11} u_7 + B_7 u_{11} u_6 u_4 + B_8 u_{11} u_5^2+
B_9 u_{11} u_5 u_4^2 +B_{10} u_{11} u_4^4 + B_{11} u_{10}u_8
\nonumber \\
& & + B_{12} u_{10} u_7 u_4 +B_{13}u_{10}u_6 u_5 +B_{14} u_{10}
u_6 u_4^2 + B_{15} u_{10} u_5^2 u_4 + B_{16} u_{10} u_5 u_4^3
\nonumber \\
& & + B_{17} u_{10} u_4^5 + B_{18} u_9^2 + B_{19} u_9 u_8 u_4+
B_{20} u_9 u_7 u_5 + B_{21} u_9 u_7 u_4^2 + B_{22} u_9 u_6^2
\nonumber \\
& & + B_{23} u_9 u_6 u_5 u_4 + B_{24} u_9 u_6 u_4^3 + B_{25}
u_9 u_5^3 + B_{26} u_9 u_5^2 u_4^2 + B_{27} u_9 u_5 u_4^4 +
B_{28}u_9 u_4^6 \nonumber \\
& & + B_{29} u_8^2 u_5  + B_{30} u_8^2 u_4^2 +B_{31}u_8 u_7 u_6
u_4 + B_{32} u_8 u_7 u_5 u_4 + B_{33} u_8 u_7 u_4^3 + B_{34}
u_8 u_6^2 u_4 \nonumber \\
& & + B_{35} u_8 u_6 u_5^2 + B_{36} u_8 u_6 u_5 u_4^2 + B_{37}
u_8 u_6 u_4^4 + B_{38} u_8 u_5^3 u_4 + B_{39} u_8 u_5^2 u_4^3 +
B_{40} u_8 u_5 u_4^5 \nonumber \\
& & + B_{41} u_8 u_4^7 +B_{42} u_7^3 +B_{43} u_7^2 u_6 u_4 +
B_{44} u_7^2 u_5^2 +B_{45}u_7^2 u_5 u_4^2 + B_{46} u_7^2 u_4^4
+ B_{47} u_7 u_6^2 u_5 \nonumber \\
& & + B_{48}u_7 u_6^2 u_4^2 + B_{49} u_7 u_6 u_5^2 u_4 + B_{50}
u_7 u_6 u_5 u_4^3+ B_{51} u_7 u_6 u_4^5+ B_{52}u_7 u_5^4
\nonumber \\
& & + B_{53} u_7 u_5^3 u_4^2 + B_{54} u_7 u_5^2 u_4^4 + B_{55}
u_7 u_5 u_4^6 + B_{56} u_7 u_4^8 + B_{57}
u_6^4 + B_{58} u_6^3 u_5 u_4  \nonumber \\
& & + B_{59}u_6^3 u_4^3 + B_{60} u_6^2 u_5^3 + B_{61} u_6^2
u_5^2 u_4^2 + B_{62}u_6^2 u_5 u_4^4+ B_{63} u_6^2 u_4^6 +
B_{64} u_6 u_5^4 u_4  \nonumber \\
& & + B_{65}u_6 u_5^3 u_4^3 + B_{66} u_6 u_5^2 u_4^5 + B_{67}
u_6 u_5 u_4^7 + B_{68} u_6 u_4^9 + B_{69} u_5^6 + B_{70} u_5^5
u_4^2  \nonumber \\
& & + B_{71} u_5^4 u_4^4 + B_{72} u_5^3 u_4^6 + B_{73} u_5^2
u_4^8 + B_{74} u_5 u_4^{10} + B_{75} u_4^{12} \eea

where $B_0=-45 a^{17} z_3.$

\section{Results and Discussion}
In this study, we introduced a new grading structure, that we
call the ``level grading". We applied this structure on the
algebra of polynomials generated by the derivatives $u_{k+i}$
over the coefficient ring $K^{(k)}$ of $C^{\infty}$ functions
of $u_i,$ \ \ $i=0,1,2,\dots,k,$ where $k$ is denoted as the
base level. We prove that this grading structure has the
property that the total derivative with respect to $x$ and the
integration by parts are filtered algebra maps. We also prove
that, if $u$ satisfies an evolution equation $u_t=F[u]$ and $F$
is a level homogeneous differential polynomial, then the total
derivative with respect to $t$ is also a filtered algebra map,
and the conserved density conditions are level homogeneous and
their top level part is independent of $u_j$ for $j<k.$ We
applied this ``level homogeneity" property on the
classification of integrable scalar evolution equations of
order $m\geq 7.$ We give explicit formulas for order $m=7,9$
and give the formulas for order $m=11,13,15$ in a closed form
with the explicit form of the coefficients $B_0.$ We observed
that, at all orders, $a$ satisfies the same equations given in
$(\ref{tanimlar1})$ The occurrence of the same  form for the
separant $a$ suggests strongly that these equations belong to a
hierarchy. The same form of $a$ has occurred in the
classification of fifth order equation \cite{G2012}, where it
has been noted that these equations would be intrinsically
related to the class of fully nonlinear third order equations
\cite{SW98},
$$u_t=F=(\alpha u_3^2+\beta u_3+\gamma )^{-1/2} \left( 2 \alpha u_3 +\beta\right)+ \delta,$$
In this equation when we compute $\frac{\partial F}{\partial u_3}$ we find that
$$a=\left(\frac{\partial F}{\partial u_3}\right)^{1/3}=\frac{1}{2} (\alpha u_3^2+\beta u_3+\gamma )^{-1/2}
(4\alpha \gamma -\beta^2)
$$
This result suggest that the equations for which we have determined the top level part belong probably to a hierarchy starting at
the fully nonlinear third order equation and the hierarchy is possibly generated by a second order recursion operator.

\newpage

\appendix
\small{ \noindent  {\bf Appendix A} \vskip 12pt

\noindent The submodules $M_i^{(k)}$ and their generating
monomials where: $i=1,2,3,...,13$ \ \ and \ \
$k=m-3,m-4,\dots,3$ used in classification of $m=7th, m=9th,
m=11th$ order evolution equations:

\noindent \textbf{Submodules with base level $k$} \bea
\hspace{-0.5cm}M_1^{(k)}&=&\langle u_{k+1}\rangle \nonumber \\
\hspace{-0.5cm}M_2^{(k)}&=&\langle u_{k+2},u_{k+1}^2\rangle \nonumber \\
\hspace{-0.5cm}M_3^{(k)}&=&\langle u_{k+3},u_{k+2}u_{k+1},u_{k+1}^3\rangle \nonumber \\
\hspace{-0.5cm}M_4^{(k)}&=&\langle u_{k+4},u_{k+3}u_{k+1},u_{k+2}^2,u_{k+2}u_{k+1}^2,u_{k+1}^4\rangle \nonumber \\
\hspace{-0.5cm}M_5^{(k)}&=&\langle u_{k+5},u_{k+4}u_{k+1},u_{k+3}u_{k+2},u_{k+3}u_{k+1}^2,u_{k+2}^2u_{k+1},u_{k+2}u_{k+1}^3,u_{k+1}^5 \rangle \nonumber \\
\hspace{-0.5cm}M_6^{(k)}&=&\langle u_{k+6},u_{k+5}u_{k+1},u_{k+4}u_{k+2},u_{k+4}u_{k+1}^2,u_{k+3}^2,u_{k+3}u_{k+2}u_{k+1},u_{k+3}u_{k+1}^3,\nonumber \\
& & u_{k+2}^3,u_{k+2}^2u_{k+1}^2,u_{k+2}u_{k+1}^4,u_{k+1}^6\rangle \nonumber \\
\hspace{-0.5cm}M_7^{(k)}&=&\langle
u_{k+7},u_{k+6}u_{k+1},u_{k+5}u_{k+2},u_{k+5}u_{k+1}^2,u_{k+4}u_{k+3},u_{k+4}u_{k+2}u_{k+1},u_{k+4}u_{k+1}^3,\nonumber
\\
& & u_{k+3}^2u_{k+1},u_{k+3}u_{k+2}^2,u_{k+3}u_{k+2}u_{k+1}^2,u_{k+3}u_{k+1}^4,\nonumber \\
& & u_{k+2}^3u_{k+1},u_{k+2}^2u_{k+1}^3,u_{k+2}u_{k+1}^5,u_{k+1}^7\rangle \nonumber \\
\hspace{-0.5cm}M_8^{(k)}&=&\langle
u_{k+8},u_{k+7}u_{k+1},u_{k+6}u_{k+2},u_{k+6}u_{k+1}^2,u_{k+5}u_{k+3},u_{k+5}u_{k+2}u_{k+1},u_{k+5}u_{k+1}^3,\nonumber\\
& & u_{k+4}^2,u_{k+4}u_{k+3}u_{k+1},u_{k+4}u_{k+2}^2,u_{k+4}u_{k+2}u_{k+1}^2,u_{k+4}u_{k+1}^4,\nonumber \\
& & u_{k+3}^2u_{k+2},u_{k+3}^2u_{k+1}^2,u_{k+3}u_{k+2}^2u_{k+1},u_{k+3}u_{k+2}u_{k+1}^3,u_{k+3}u_{k+1}^5,\nonumber \\
& & u_{k+2}^4,u_{k+2}^3u_{k+1}^2,u_{k+2}^2u_{k+1}^4,u_{k+2}u_{k+1}^6,u_{k+1}^8\rangle \nonumber \\
\hspace{-0.5cm}M_9^{(k)}&=&\langle
u_{k+9},u_{k+8}u_{k+1},u_{k+7}u_{k+2},u_{k+7}u_{k+1}^2,\nonumber \\
& & u_{k+6}u_{k+3},u_{k+6}u_{k+2}u_{k+1},u_{k+6}u_{k+1}^3,\nonumber \\
& & u_{k+5}u_{k+4},u_{k+5}u_{k+3}u_{k+1},u_{k+5}u_{k+2}^2,u_{k+5}u_{k+2}u_{k+1}^2,u_{k+5}u_{k+1}^4,\nonumber\\
& & u_{k+4}^2u_{k+1},u_{k+4}u_{k+3}u_{k+2},u_{k+4}u_{k+3}u_{k+1}^2,u_{k+4}u_{k+2}^2u_{k+1},u_{k+4}u_{k+2}u_{k+1}^3,u_{k+4}u_{k+1}^5,\nonumber \\
& & u_{k+3}^3,u_{k+3}^2u_{k+2}u_{k+1},u_{k+3}^2u_{k+1}^3,u_{k+3}u_{k+2}^3,u_{k+3}u_{k+2}^2u_{k+1}^2,u_{k+3}u_{k+2}u_{k+1}^4,u_{k+3}u_{k+1}^6,\nonumber \\
& & u_{k+2}^4u_{k+1},u_{k+2}^3u_{k+1}^3,u_{k+2}^2u_{k+1}^5,u_{k+2}u_{k+1}^7,u_{k+1}^9 \rangle \nonumber \\
\hspace{-0.5cm}M_{10}^{(k)}&=&\langle u_{k+10},u_{k+9}u_{k+1},u_{k+8}u_{k+2},u_{k+8}u_{k+1}^2,\nonumber \\
& & u_{k+7}u_{k+3},u_{k+7}u_{k+2}u_{k+1},u_{k+7}u_{k+1}^3,\nonumber \\
& & u_{k+6}u_{k+4},u_{k+6}u_{k+3}u_{k+1},u_{k+6}u_{k+2}^2,u_{k+6}u_{k+2}u_{k+1}^2,u_{k+6}u_{k+1}^4,\nonumber \\
& & u_{k+5}^2,u_{k+5}u_{k+4}u_{k+1},u_{k+5}u_{k+3}u_{k+2},u_{k+5}u_{k+3}u_{k+1}^2,u_{k+5}u_{k+2}^2u_{k+1},u_{k+5}u_{k+2}u_{k+1}^3,u_{k+5}u_{k+1}^5,\nonumber \\
& & u_{k+4}^2u_{k+2},u_{k+4}^2u_{k+1}^2,u_{k+4}u_{k+3}^2,u_{k+4}u_{k+3}u_{k+2}u_{k+1},u_{k+4}u_{k+3}u_{k+1}^3,\nonumber \\
& & u_{k+4}u_{k+2}^3,u_{k+4}u_{k+2}^2u_{k+1}^2,u_{k+4}u_{k+2}u_{k+1}^4,u_{k+4}u_{k+1}^6,\nonumber \\
& & u_{k+3}^3u_{k+1},u_{k+3}^2u_{k+2}^2,u_{k+3}^2u_{k+2}u_{k+1}^2,u_{k+3}^2u_{k+1}^4,u_{k+3}u_{k+2}^3u_{k+1},u_{k+3}u_{k+2}^2u_{k+1}^3,u_{k+3}u_{k+2}u_{k+1}^5,u_{k+3}u_{k+1}^7,\nonumber \\
& & u_{k+2}^5,u_{k+2}^4u_{k+1}^2,u_{k+2}^3u_{k+1}^4,u_{k+2}^2u_{k+1}^6,u_{k+2}u_{k+1}^8,u_{k+1}^{10}\rangle \nonumber \\
\hspace{-0.5cm}M_{11}^{(k)}&=&\langle
u_{k+11},u_{k+10}u_{k+1},u_{k+9}u_{k+2},u_{k+9}u_{k+1}^2,\nonumber \\
& & u_{k+8}u_{k+3},u_{k+8}u_{k+2}u_{k+1},u_{k+8}u_{k+1}^3,\nonumber \\
& & u_{k+7}u_{k+4},u_{k+7}u_{k+3}u_{k+1},u_{k+7}u_{k+2}^2,u_{k+7}u_{k+2}u_{k+1}^2,u_{k+7}u_{k+1}^4,\nonumber \\
& & u_{k+6}u_{k+5},u_{k+6}u_{k+4}u_{k+1},u_{k+6}u_{k+3}u_{k+2},u_{k+6}u_{k+3}u_{k+1}^2,u_{k+6}u_{k+2}^2u_{k+1},u_{k+6}u_{k+2}u_{k+1}^3,u_{k+6}u_{k+1}^5\nonumber \\
& & u_{k+5}^2u_{k+1},u_{k+5}u_{k+4}u_{k+2},u_{k+5}u_{k+4}u_{k+1}^2,u_{k+5}u_{k+3}^2,u_{k+5}u_{k+3}u_{k+2}u_{k+1},u_{k+5}u_{k+3}u_{k+1}^3,\nonumber \\
& & u_{k+5}u_{k+2}^3,u_{k+5}u_{k+2}^2u_{k+1}^2,u_{k+5}u_{k+2}u_{k+1}^4, u_{k+5}u_{k+1}^6,\nonumber \\
& & u_{k+4}^2u_{k+3},u_{k+4}^2u_{k+2}u_{k+1},u_{k+4}^2u_{k+1}^3,\nonumber \\
& & u_{k+4}u_{k+3}^2u_{k+1},u_{k+4}u_{k+3}u_{k+2}^2,u_{k+4}u_{k+3}u_{k+2}u_{k+1}^2,u_{k+4}u_{k+3}u_{k+1}^4,\nonumber \\
& & u_{k+4}u_{k+2}^3u_{k+1},u_{k+4}u_{k+2}^2u_{k+1}^3,u_{k+4}u_{k+2}u_{k+1}^5,u_{k+4}u_{k+1}^7,\nonumber \\
& &
u_{k+3}^3u_{k+2},u_{k+3}^3u_{k+1}^2,u_{k+3}^2u_{k+2}^2u_{k+1},u_{k+3}^2u_{k+2}u_{k+1}^3,u_{k+3}^2u_{k+1}^5,\nonumber
\\
& & u_{k+3}u_{k+2}^4,u_{k+3}u_{k+2}^3u_{k+1}^2,u_{k+3}u_{k+2}^2u_{k+1}^4,u_{k+3}u_{k+2}u_{k+1}^6,u_{k+3}u_{k+1}^8,\nonumber \\
& & u_{k+2}^5u_{k+1},u_{k+3}^4u_{k+1}^3,u_{k+2}^3u_{k+1}^5,u_{k+2}^2u_{k+1}^7,u_{k+2}u_{k+1}^9,u_{k+1}^{11}\rangle \nonumber \\
\hspace{-0.5cm}M_{12}^{(k)}&=&\langle u_{k+12},u_{k+11}u_{k+1},u_{k+10}u_{k+2},u_{k+10}u_{k+1}^2,u_{k+9}u_{k+3},u_{k+9}u_{k+2}u_{k+1},u_{k+9}u_{k+1}^3,\nonumber \\
& & u_{k+8}u_{k+4},u_{k+8}u_{k+3}u_{k+1},u_{k+8}u_{k+2}^2,u_{k+8}u_{k+2}u_{k+1}^2,u_{k+8}u_{k+1}^4, \nonumber \\
& & u_{k+7}u_{k+5},u_{k+7}u_{k+4}u_{k+1},u_{k+7}u_{k+3}u_{k+2},u_{k+7}u_{k+3}u_{k+1}^2,u_{k+7}u_{k+2}^2u_{k+1},u_{k+7}u_{k+2}u_{k+1}^3,u_{k+7}u_{k+1}^5,\nonumber \\
& & u_{k+6}^2,u_{k+6}u_{k+5}u_{k+1},u_{k+6}u_{k+4}u_{k+2},u_{k+6}u_{k+4}u_{k+1}^2,u_{k+6}u_{k+3}^2,u_{k+6}u_{k+3}u_{k+2}u_{k+1},u_{k+6}u_{k+3}u_{k+1}^3,\nonumber \\
& & u_{k+6}u_{k+2}^3,u_{k+6}u_{k+2}^2u_{k+1}^2,u_{k+6}u_{k+2}u_{k+1}^4,u_{k+6}u_{k+1}^6,\nonumber\\
& & u_{k+5}^2u_{k+2},u_{k+5}^2u_{k+1}^2,u_{k+5}u_{k+4}u_{k+3},u_{k+5}u_{k+4}u_{k+2}u_{k+1},u_{k+5}u_{k+4}u_{k+1}^3,u_{k+5}u_{k+3}^2u_{k+1},\nonumber \\
& & u_{k+5}u_{k+3}u_{k+2}^2,u_{k+5}u_{k+3}u_{k+2}u_{k+1}^2,u_{k+5}u_{k+3}u_{k+1}^4,u_{k+5}u_{k+2}^3u_{k+1},u_{k+5}u_{k+2}^2u_{k+1}^3,u_{k+5}u_{k+2}u_{k+1}^5,u_{k+5}u_{k+1}^7,\nonumber \\
& & u_{k+4}^3,u_{k+4}^2u_{k+3}u_{k+1},u_{k+4}^2u_{k+2}^2,u_{k+4}^2u_{k+2}u_{k+1}^2,u_{k+4}^2u_{k+1}^4,\nonumber \\
& & u_{k+4}u_{k+3}^2u_{k+2},u_{k+4}u_{k+3}^2u_{k+1}^2,u_{k+4}u_{k+3}u_{k+2}^2u_{k+1},u_{k+4}u_{k+3}u_{k+2}u_{k+1}^3,u_{k+4}u_{k+3}u_{k+1}^5,\nonumber \\
& & u_{k+4}u_{k+2}^4,u_{k+4}u_{k+2}^3u_{k+1}^2,u_{k+4}u_{k+2}^2u_{k+1}^4,u_{k+4}u_{k+2}u_{k+1}^6,u_{k+4}u_{k+1}^8,\nonumber \\
& & u_{k+3}^4,u_{k+3}^3u_{k+2}u_{k+1},u_{k+3}^3u_{k+1}^3,u_{k+3}^2u_{k+2}^3,u_{k+3}^2u_{k+2}^2u_{k+1}^2,u_{k+3}^2u_{k+2}u_{k+1}^4,u_{k+3}^2u_{k+1}^6,\nonumber \\
& & u_{k+3}u_{k+2}^4u_{k+1},u_{k+3}u_{k+2}^3u_{k+1}^3,u_{k+3}u_{k+2}^2u_{k+1}^5,u_{k+3}u_{k+2}u_{k+1}^7,u_{k+3}u_{k+1}^9,\nonumber \\
& & u_{k+2}^6,u_{k+2}^5u_{k+1}^2,u_{k+2}^4u_{k+1}^4,u_{k+2}^3u_{k+1}^6,u_{k+2}^2u_{k+1}^8,u_{k+2}u_{k+1}^{10},u_{k+1}^{12}\rangle \nonumber \\
\hspace{-0.5cm}M_{13}^{(k)}&=&\langle u_{k+13},u_{k+12}u_{k+1},u_{k+11}u_{k+2},u_{k+11}u_{k+1}^2,u_{k+10}u_{k+3},u_{k+10}u_{k+2}u_{k+1},u_{k+10}u_{k+1}^3,\nonumber\\
& & u_{k+9}u_{k+4},u_{k+9}u_{k+3}u_{k+1},u_{k+9}u_{k+2}^2,u_{k+9}u_{k+2}u_{k+1}^2,u_{k+9}u_{k+1}^4,\nonumber \\
& & u_{k+8}u_{k+5},u_{k+8}u_{k+4}u_{k+1},u_{k+8}u_{k+3}u_{k+2},u_{k+8}u_{k+3}u_{k+1}^2,u_{k+8}u_{k+2}^2u_{k+1},u_{k+8}u_{k+2}u_{k+1}^3,u_{k+8}u_{k+1}^5,\nonumber \\
& &
u_{k+7}u_{k+6},u_{k+7}u_{k+5}u_{k+1},u_{k+7}u_{k+4}u_{k+2},u_{k+7}u_{k+4}u_{k+1}^2,u_{k+7}u_{k+3}^2,u_{k+7}u_{k+3}u_{k+2}u_{k+1},u_{k+7}u_{k+3}u_{k+1}^3,\nonumber
\\
& & u_{k+7}u_{k+2}^3,u_{k+7}u_{k+2}^2u_{k+1}^2,u_{k+7}u_{k+2}u_{k+1}^4,u_{k+7}u_{k+1}^6,u_{k+6}^2u_{k+1},u_{k+6}u_{k+5}u_{k+2},u_{k+6}u_{k+5}u_{k+1}^2,\nonumber \\
& & u_{k+6}u_{k+4}u_{k+3},u_{k+6}u_{k+4}u_{k+2}u_{k+1},u_{k+6}u_{k+4}u_{k+1}^3,\nonumber \\
& & u_{k+6}u_{k+3}^2u_{k+1},u_{k+6}u_{k+3}u_{k+2}^2,u_{k+6}u_{^+3}u_{k+2}u_{k+1}^2,u_{k+6}u_{k+3}u_{k+1}^4,\nonumber \\
& & u_{k+6}u_{k+2}^3u_{k+1},u_{k+6}u_{k+2}^2u_{k+1}^3,u_{k+6}u_{k+2}u_{k+1}^5,u_{k+6}u_{k+1}^7,\nonumber \\
& & u_{k+5}^2u_{k+3},u_{k+5}^2u_{k+2}u_{k+1},u_{k+5}^2u_{k+1}^3,u_{k+5}u_{k+4}^2,\nonumber \\
& & u_{k+5}u_{k+4}u_{k+3}u_{k+1},u_{k+5}u_{k+4}u_{k+2}^2,u_{k+5}u_{k+4}u_{k+2}u_{k+1}^2,u_{k+5}u_{k+4}u_{k+1}^4,\nonumber \\
& & u_{k+5}u_{k+3}^2u_{k+2},u_{k+5}u_{k+3}^2u_{k+1}^2,u_{k+5}u_{k+3}u_{k+2}^2u_{k+1},u_{k+5}u_{k+3}u_{k+2}u_{k+1}^3,u_{k+5}u_{k+3}u_{k+1}^5,\nonumber \\
& & u_{k+5}u_{k+2}^4,u_{k+5}u_{k+2}^3u_{k+1}^2,u_{k+5}u_{k+2}^2u_{k+1}^4,u_{k+5}u_{k+2}u_{k+1}^6,u_{k+5}u_{k+1}^8,\nonumber \\
& & u_{k+4}^3u_{k+1},u_{k+4}^2u_{k+3}u_{k+2},u_{k+4}^2u_{k+3}u_{k+1}^2,u_{k+4}^2u_{k+2}^2u_{k+1},u_{k+4}^2u_{k+2}u_{k+1}^3,u_{k+4}^2u_{k+1}^5,\nonumber \\
& & u_{k+4}u_{k+3}^3,u_{k+4}u_{k+3}^2u_{k+2}u_{k+1},u_{k+4}u_{k+3}^2u_{k+1}^3,\nonumber \\
& & u_{k+4}u_{k+3}u_{k+2}^3,u_{k+4}u_{k+3}u_{k+2}^2u_{k+1}^2,u_{k+4}u_{k+3}u_{k+2}u_{k+1}^4,u_{k+4}u_{k+3}u_{k+1}^6,\nonumber \\
& & u_{k+4}u_{k+2}^4u_{k+1},u_{k+4}u_{k+2}^3u_{k+1}^3,u_{k+4}u_{k+2}^2u_{k+1}^5,u_{k+4}u_{k+2}u_{k+1}^7,u_{k+4}u_{k+1}^9,\nonumber \\
& & u_{k+3}^4u_{k+1},u_{k+3}^3u_{k+2}^2,u_{k+3}^3u_{k+2}u_{k+1}^2,u_{k+3}^3u_{k+1}^4,u_{k+3}^2u_{k+2}^3u_{k+1},u_{k+3}^2u_{k+2}^2u_{k+1}^3,u_{k+3}^2u_{k+2}u_{k+1}^5,u_{k+3}^2u_{k+1}7^7,\nonumber \\
& & u_{k+3}u_{k+2}^5,u_{k+3}u_{k+2}^4u_{k+1}^2,u_{k+3}u_{k+2}^3u_{k+1}^4,u_{k+3}u_{k+2}^2u_{k+1}^6,u_{k+3}u_{k+2}u_{k+1}^8,u_{k+3}u_{k+1}^{10},\nonumber \\
& &
u_{k+2}^6u_{k+1},u_{k+2}^5u_{k+1}^3,u_{k+2}^4u_{k+1}^5,u_{k+2}^3u_{k+1}^7,u_{k+2}^2u_{k+1}^9,u_{k+2}u_{k+1}^{11},u_{k+1}^{13}\rangle
\nonumber \eea \vspace*{0.6cm}

\appendix
\vskip 12pt
\noindent {\bf Appendix B}\\ \\
The quotient submodules $\overline{M_i^{(k)}}$ and their
generating monomials (that are not total derivatives), where
$k=m-3,m-4,\dots,3$ and $i=1,2,3,...,11,13$ used in
classification of $m=7th, m=9th, m=11th$ order evolution
equations:

\noindent \textbf{Quotient Submodules with base level $k$}
 \bea
 \overline{M_1^{(k)}}&=& \langle \emptyset \rangle \nonumber  \\
  \overline{M_2^{(k)}}&=&\langle u_{k+1}^2 \rangle \nonumber \\
 \overline{M_3^{(k)}}&=&\langle u_{k+1}^3 \rangle \nonumber \\
 \overline{M_4^{(k)}}&=&\langle u_{k+2}^2,u_{k+1}^4 \rangle \nonumber \\
 \overline{M_5^{(k)}}&=&\langle u_{k+2}^2u_{k+1},u_{k+1}^5 \rangle \nonumber \\
 \overline{M_6^{(k)}}&=&\langle u_{k+3}^2,u_{k+2}^3,u_{k+2}^2u_{k+1}^2,u_{k+1}^6 \rangle \nonumber \\
 \overline{M_7^{(k)}}&=&\langle u_{k+3}^2u_{k+1},u_{k+2}^3u_{k+1},u_{k+2}^2u_{k+1}^3,u_{k+1}^7 \rangle \nonumber \\
\overline{M_8^{(k)}}&=&\langle u_{k+4}^2,u_{k+3}^2u_{k+2},u_{k+3}^2u_{k+1}^2,u_{k+2}^4,u_{k+2}^3u_{k+1}^2,u_{k+2}^2u_{k+1}^4,u_{k+1}^8 \rangle \nonumber \\
\overline{M_9^{(k)}}&=&\langle u_{k+4}^2u_{k+1},u_{k+3}^3,u_{k+3}^2u_{k+2}u_{k+1},u_{k+3}^2u_{k+1}^3,u_{k+2}^4u_{k+1},u_{k+2}^3u_{k+1}^3,u_{k+2}^2u_{k+1}^5,u_{k+1}^9 \rangle \nonumber  \\
 \overline{M_{10}^{(k)}}&=&\langle u_{k+5}^2,u_{k+4}^2u_{k+2},u_{k+4}^2u_{k+1}^2,u_{k+3}^3u_{k+1},u_{k+3}^2u_{k+2}^2,u_{k+3}^2u_{k+2}u_{k+1}^2,u_{k+3}^2u_{k+1}^4,\nonumber \\
& & u_{k+2}^5,u_{k+2}^4u_{k+1}^2,u_{k+2}^3u_{k+1}^4,u_{k+2}^2u_{k+1}^6,u_{k+1}^{10}\rangle \nonumber \\
 \overline{M_{11}^{(k)}}&=&\langle
u_{k+5}^2u_{k+1},u_{k+4}^2u_{k+3},u_{k+4}^2u_{k+2}u_{k+1},u_{k+4}^2u_{k+1}^3,\nonumber \\
& & u_{k+3}^3u_{k+2},u_{k+3}^3u_{k+1}^2,u_{k+3}^2u_{k+2}^2u_{k+1},u_{k+3}^2u_{k+2}u_{k+1}^3,u_{k+2}^5u_{k+1},u_{k+3}^2u_{k+1}^5, \nonumber \\
& & u_{k+2}^4u_{k+1}^3,u_{k+2}^3u_{k+1}^5,u_{k+2}^2u_{k+1}^7,u_{k+1}^{11}\rangle \nonumber \\
 \overline{M_{12}^{(k)}}&=&\langle
u_{k+6}^2,u_{k+5}^2u_{k+2},u_{k+5}^2u_{k+1}^2,u_{k+4}^3,u_{k+4}^2u_{k+3}u_{k+1},u_{k+4}^2u_{k+2}^2,u_{k+4}^2u_{k+2}u_{k+1}^2,u_{k+4}^2u_{k+1}^4,\nonumber \\
& & u_{k+3}^4,u_{k+3}^3u_{k+2}u_{k+1},u_{k+3}^3u_{k+1}^3,u_{k+3}^2u_{k+2}^3,u_{k+3}^2u_{k+2}^2u_{k+1}^2,u_{k+3}^2u_{k+2}u_{k+1}^4,u_{k+3}^2u_{k+1}^6,\nonumber \\
& & u_{k+2}^6,u_{k+2}^5u_{k+1}^2,u_{k+2}^4u_{k+1}^4,u_{k+2}^3u_{k+1}^6,u_{k+2}^2u_{k+1}^8,u_{k+1}^{12}\rangle \nonumber \\
 \overline{M_{13}^{(k)}}&=&\langle
u_{k+6}^2u_{k+1},u_{k+5}^2u_{k+3},u_{k+5}^2u_{k+2}u_{k+1},u_{k+5}^2u_{k+1}^3,\nonumber \\
& & u_{k+4}^3u_{k+1},u_{k+4}^2u_{k+3}u_{k+2},u_{k+4}^2u_{k+3}u_{k+1}^2,u_{k+4}^2u_{k+2}^2u_{k+1},u_{k+4}^2u_{k+2}u_{k+1}^3, \nonumber \\
& & u_{k+3}^4u_{k+1},u_{k+3}^3u_{k+2}^2,u_{k+3}^3u_{k+2}u_{k+1}^2,u_{k+3}^3u_{k+1}^4,u_{k+3}^2u_{k+2}^3u_{k+1},u_{k+3}^2u_{k+2}^2u_{k+1}^3,u_{k+3}^2u_{k+2}u_{k+1}^5,u_{k+3}^2u_{k+1}^7,\nonumber \\
& &
u_{k+2}^6u_{k+1},u_{k+2}^5,u_{k+1}^3,u_{k+2}^4u_{k+1}^5,u_{k+2}^3u_{k+1}^7,u_{k+2}^2u_{k+1}^9,u_{k+1}^{13}
\rangle \nonumber \eea

\end{document}